\documentclass[superscriptaddress,onecolumn,secnumarabic,
amssymb,amsmath,nobibnotes,aps,prd,showkeys,showpacs,nofootinbib]{revtex4}

\usepackage[latin1]{inputenc}
\usepackage{graphicx}
\usepackage[english]{babel}

\usepackage{amsmath}
\usepackage{amssymb}
\usepackage{amsfonts}
\usepackage{colordvi}
\usepackage{psfrag}
\usepackage{color}

\begin{document}

\def\cL{{\cal L}}
\def\be{\begin{equation}}
\def\ee{\end{equation}}
\def\bea{\begin{eqnarray}}
\def\eea{\end{eqnarray}}
\def\beq{\begin{eqnarray}}
\def\eeq{\end{eqnarray}}
\def\tr{{\rm tr}\, }
\def\nn{\nonumber \\}
\def\e{{\rm e}}


\def\bef{\begin{figure}}
\def\eef{\end{figure}}
\newcommand{\ans}{ansatz }
\newcommand{\eeqn}{\end{eqnarray}}
\newcommand{\bd}{\begin{displaymath}}
\newcommand{\ed}{\end{displaymath}}
\newcommand{\mat}[4]{\left(\begin{array}{cc}{#1}&{#2}\\{#3}&{#4}
\end{array}\right)}
\newcommand{\matr}[9]{\left(\begin{array}{ccc}{#1}&{#2}&{#3}\\
{#4}&{#5}&{#6}\\{#7}&{#8}&{#9}\end{array}\right)}
\newcommand{\matrr}[6]{\left(\begin{array}{cc}{#1}&{#2}\\
{#3}&{#4}\\{#5}&{#6}\end{array}\right)}
\newcommand{\cvb}[3]{#1^{#2}_{#3}}
\def\lsim{\raise0.3ex\hbox{$\;<$\kern-0.75em\raise-1.1ex
e\hbox{$\sim\;$}}}
\def\gsim{\raise0.3ex\hbox{$\;>$\kern-0.75em\raise-1.1ex
\hbox{$\sim\;$}}}
\def\abs#1{\left| #1\right|}
\def\simlt{\mathrel{\lower2.5pt\vbox{\lineskip=0pt\baselineskip=0pt
           \hbox{$<$}\hbox{$\sim$}}}}
\def\simgt{\mathrel{\lower2.5pt\vbox{\lineskip=0pt\baselineskip=0pt
           \hbox{$>$}\hbox{$\sim$}}}}
\def\unity{{\hbox{1\kern-.8mm l}}}
\newcommand{\eps}{\varepsilon}
\def\ep{\epsilon}
\def\ga{\gamma}
\def\Ga{\Gamma}
\def\om{\omega}
\def\omp{{\omega^\prime}}
\def\Om{\Omega}
\def\la{\lambda}
\def\La{\Lambda}
\def\al{\alpha}
\newcommand{\ov}{\overline}
\renewcommand{\to}{\rightarrow}
\renewcommand{\vec}[1]{\mathbf{#1}}
\newcommand{\vect}[1]{\mbox{\boldmath$#1$}}
\def\tm{{\widetilde{m}}}
\def\mcirc{{\stackrel{o}{m}}}
\newcommand{\Dm}{\Delta m}
\newcommand{\dm}{\varepsilon}
\newcommand{\tanb}{\tan\beta}
\newcommand{\nbar}{\tilde{n}}
\newcommand\PM[1]{\begin{pmatrix}#1\end{pmatrix}}
\newcommand{\up}{\uparrow}
\newcommand{\down}{\downarrow}
\def\omE{\omega_{\rm Ter}}
%

\newcommand{\Dsusy}{{susy \hspace{-9.4pt} \slash}\;}
\newcommand{\DCP}{{CP \hspace{-7.4pt} \slash}\;}
\newcommand{\mc}{\mathcal}
\newcommand{\gr}{\mathbf}
\renewcommand{\to}{\rightarrow}
\newcommand{\gtc}{\mathfrak}
\newcommand{\wh}{\widehat}
\newcommand{\br}{\langle}
\newcommand{\kt}{\rangle}


\def\lsim{\mathrel{\mathop  {\hbox{\lower0.5ex\hbox{$\sim$}
\kern-0.8em\lower-0.7ex\hbox{$<$}}}}}
\def\gsim{\mathrel{\mathop  {\hbox{\lower0.5ex\hbox{$\sim$}
\kern-0.8em\lower-0.7ex\hbox{$>$}}}}}

\def\nn{\\  \nonumber}
\def\de{\partial}
\def\brf{{\mathbf f}}
\def\bbf{\bar{\bf f}}
\def\bF{{\bf F}}
\def\bbF{\bar{\bf F}}
\def\bA{{\mathbf A}}
\def\bB{{\mathbf B}}
\def\bG{{\mathbf G}}
\def\bI{{\mathbf I}}
\def\bM{{\mathbf M}}
\def\bY{{\mathbf Y}}
\def\bX{{\mathbf X}}
\def\bS{{\mathbf S}}
\def\bb{{\mathbf b}}
\def\bh{{\mathbf h}}
\def\bg{{\mathbf g}}
\def\bla{{\mathbf \la}}
\def\bmu{\mathbf m }
\def\by{{\mathbf y}}
\def\bmu{\mbox{\boldmath $\mu$} }
\def\bsig{\mbox{\boldmath $\sigma$} }
\def\bunity{{\mathbf 1}}
\def\cA{{\cal A}}
\def\cB{{\cal B}}
\def\cC{{\cal C}}
\def\cD{{\cal D}}
\def\cF{{\cal F}}
\def\cG{{\cal G}}
\def\cH{{\cal H}}
\def\cI{{\cal I}}
\def\cL{{\cal L}}
\def\cN{{\cal N}}
\def\cM{{\cal M}}
\def\cO{{\cal O}}
\def\cR{{\cal R}}
\def\cS{{\cal S}}
\def\cT{{\cal T}}
\def\eV{{\rm eV}}

\title{Suppression of Bekenstein-Hawking radiation in $f(T)$-gravity}

\author{Andrea Addazi}

\affiliation{ Center for Field Theory and Particle Physics \& Department of Physics, Fudan University, 200433 Shanghai, China}


\date{\today}

\begin{abstract}

We discuss semiclassical Nariai black holes
in the framework of $f(T)$-gravity.
For a diagonal choice of tetrades, 
stable Nariai metrics can be found, emitting 
Bekenstein-Hawking radiation in semiclassical limit. 
However, for a non-diagonal choice of tetrades, 
evaporation and antievaporation instabilities
are turned on. 
In turn, this causes a back-reaction effect 
suppressing the Bekenstein-Hawking radiation. 
In particular, evaporation instabilities 
produce 
a new radiation -- different by Bekenstein-Hawking emission --
non-violating unitarity in particle physics sector.


\end{abstract}

\maketitle

\section{Introduction}

In extended theories of gravity, extremal Schwarzschild-de Sitter black holes, 
also called Nariai solution, 
can be unstable. 
This phenomena was firstly discovered by Bousso, Hawking,
Nojiri and Odintsov in context of quantum dilaton-gravity 
\cite{Bousso:1997wi,Nojiri:1998ph,Nojiri:1998ue,Elizalde:1999dw}
and re-discovered in various extensions of General Relativity
\cite{Nojiri:2013su,Nojiri:2014jqa,Sebastiani:2013fsa,Houndjo:2013qna,Oikonomou:2016fxb,Oikonomou:2015lgy,Addazi:2016hip,Katsuragawa:2014hda,Chakraborty:2014xla,Chakraborty:2015bja,Chakraborty:2016ydo,Addazi:2017puj,Singh:2017qur} later on.

In this paper, we will consider quantum semiclassical Black holes in $f(T)$-gravity. 
We will discuss all possible classes of Narai black holes
for diagonal or non-diagonal tetrads.
In the non-diagonal tetrads case, 
evaporation and antievaporation instabilties 
of the Nariai solution appear out. 
In such a gauge choice, 
we will show that 
evaporating and antievaporating solutions turn off Bekenstein-Hawking radiation. 
Such a surprising result is based on the
trapped surface analysis developed by Ellis and Firouzjaee in
Refs. \cite{Ellis:2013oka,Firouzjaee:2014zfa,Firouzjaee:2015bqa}
\footnote{Other further analysis based on Ellis and Firouzjaee were discussed in 
Refs.\cite{Addazi:2016prb,Helou:2015yqa,Helou:2015zma,Helou:2016xyu}.}.
The pair creation process in a dynamical space-time is suppressed 
by the energy conservation and the causality condition.
 Particularly interesting is the evaporation phenomena in the non-diagonal choice.
Contrary to Bekenstein-Hawking emission, the new evaporation effect 
does not lead to any information paradox or firewalls. 
Classical evaporation will push out {\it both} Bekenstein-Hawking pair
before the effective tunneling time, 
i.e.  there will be not 
any entanglement of the interior and the external regions of the black hole. 
More specifically, 
our argument is based on the analysis of the Raychaudhuri equation in $f(T)$-gravity \cite{Cai:2015emx}
\footnote{See Refs.  \cite{Basilakos:2013rua,Paliathanasis:2014iva,Paliathanasis:2016vsw,Oikonomou:2016jjh,Oikonomou:2017isf,Paliathanasis:2017flf}
 for recent progresses on $f(T)$-gravity}. 
In the antievaporation case, the torsion contribution 
 introduces a gravitational focusing term into the  
Raychaduri equation.
As a consequence, 
an emitting marginally trapped surface 
 will transit from a time-like to a space-like manifold. 
In evaporation, on the contrary, the torsion term induces an anti-focusing contribution
in the Raychaduri equation. 

The  paper is organized as follows:
in Section 2, we briefly review basis of $f(T)$-gravity,
 Section 3-4 are devoted to a review of Nariai solutions in $f(T)$-gravity. 
In Section 5, our main argument is discussed. 
In Section 6, conclusions and further comments are shown. 

\section{$f(T)$-gravity}

The $f(T)$-action reads as follows: 
\be \label{action}
{I}=\frac{1}{16\pi}\int d^{4}x\sqrt{-g}f(T)+S_{m}\, ,
\ee
in units $G=c=1$, 
where we project the metric and coordinates in representation of the tetrad matrices
\be \label{ds}
ds^{2}=g_{\mu\nu}dx^{\mu}dx^{\nu}=\eta_{ij}\theta^{i}\theta^{j}\, ,
\ee
\be \label{dx}
dx^{\mu}=e_{i}^{\mu}\theta^{i},\,\,\,\,\,\,\,\theta^{i}=e^{i}_{\mu}dx^{\mu}\, ,
\ee
where $e_{i}^{\mu}e^{i}_{\nu}=\delta_{\nu}^{\mu}$, $\eta_{ij}={\rm diag}(-1,1,1,1)$,
$\sqrt{-g}=e={\rm det}[e_{\mu}^{i}]$, $i,j=1,2,3,4$, $\mu,\nu=0,1,2,3$.
We have introduced above the torsion tensor as follows:
\be \label{T}
T_{\mu\nu}^{\alpha}=\Gamma_{\nu\mu}^{\alpha}-\Gamma_{\mu\nu}^{\alpha}=e^{\alpha}_{j}(\partial_{\mu}e_{\nu}^{i}-\partial_{\nu}e_{\mu}^{i})\, 
\ee
The Eulero-Lagrange equations can be obtained by varying 
the action with respect to the tetrad field $e_{\mu}^{i}$:
\be \label{EOM}
S_{\mu}^{\nu\rho}\partial_{\rho}T \frac{d^{2}f}{dT^{2}}+[e^{-1}e_{\mu}^{i}\partial_{\rho}(e S_{\alpha}^{\nu\rho}e_{i}^{\alpha})+T^{\alpha}_{\mu\sigma}S^{\nu\sigma}_{\alpha}]\frac{df}{dT}+\frac{1}{2}\delta_{\mu}^{\nu}f=4\pi T^{(m)}_{\mu\nu}\, ,
\ee
where $T^{(m)}_{\mu\nu}$ is the energy-momentum tensor
and $S_{\mu}^{\nu\rho}$ 
\be \label{S}
S_{\alpha}^{\mu\nu}=\frac{1}{2}(\delta_{\alpha}^{\mu}T^{\nu\beta}_{\beta}-\delta_{\beta}^{\mu}T^{\nu\beta}_{\alpha}+K_{\alpha}^{\mu\nu})\, ,
\ee
where $K_{\alpha}^{\mu\nu}$
is the cotorsion
and the scalar torsion is 
 \be \label{S}
T=T_{\mu\nu}^{\alpha}S_{\alpha}^{\mu\nu}\, .
\ee
General relativity with a cosmological constant is recovered in the limit $\frac{d^{2}f}{dT^{2}}\rightarrow 0$, i.e. $f(T)=a + bT$

\section{Nariai Black hole in diagonal tetrads}  

The Nariai space-time has the following form:
\be \label{Narai}
ds^{2}=\frac{1}{\Lambda}\left[-\frac{1}{\cos^{2}\tau}(dx^{2}-d\tau^{2})+d\Omega^{2}\right]\,,
\ee
where $\Lambda$ is the cosmological constant, and the $d\Omega^{2}$ is the 
solid angle on a 2-sphere $d\Omega^{2}=d\theta^{2}+\sin^{2}\theta d\psi^{2}$, 
$0<\tau<\pi/2$, $0<t<\infty$, $\cosh t=1/\cos \tau$.
The Ricci scalar of Nariai space-time is 
 a constant being  $R=4\Lambda$.
 The Nariai space-time is a solution of Eq.(\ref{EOM})
 in the diagonal tetrad ansatz 
 \be \label{pertNarai}
ds^{2}=e^{2\rho(x,t)}(-dx^{2}+d\tau^{2})+e^{-2\phi(x,t)}d\Omega^{2}\, ,
\ee
\be \label{pertNarai2}
e_{\mu}^{a}=[ e^{\rho},e^{\rho},e^{-\phi},e^{-\phi}\sin \theta]\, .
\ee

Dynamical aspects of Nariai solutions can be studied 
with pertubation theory methods: 
 \be \label{pertNarai3}
\rho=-ln[\sqrt{\Lambda}\cos \tau]+\delta \rho(\tau,x)
\ee
\be \label{pertNarai4}
\phi={\rm ln}\sqrt{\Lambda}+\delta \phi(\tau,x)
\ee
and 
\be \label{pertNarai5}
\delta T=-2\Lambda \sin(2\tau)\delta \dot{\phi}
\ee
is found. 
Inserting Eq.(\ref{pertNarai4},\ref{pertNarai5}) in Eq.(\ref{EOM})
one finds
\be \label{pertNarai6}
\delta \phi(x,\tau)=k_{1}\sin(x-\bar{x}) \sec \tau+k_{2}
\ee
where $\bar{x}$ is the fixed initial condition, 
$k_{1,2,}$ are two integration constants. 
Now, the horizon is defined through the condition 
\be \label{pertNarai7}
\left(\frac{\partial \delta \phi}{\partial \tau}\right)^{2}=\left(\frac{\partial \delta \phi}{\partial x}\right)^{2} \, .
\ee
From this, we obtain 
\be \label{pertNaraia8}
x_{h}=\bar{x}-\tau+m\pi-\frac{\pi}{2}\, ,
\ee
where $m=0,1,...$, 
corresponding 
to 
\be \label{pertNaraib}
\delta\phi(\tau,x_{h})=k_{1}(-1)^{n+1}+k_{2}\, ,
\ee
\be \label{pertNaraic}
r_{h}(\tau)^{-2}=1+ \delta\phi(\tau,x_{h})\, .
\ee
This means that the Black hole radius is fixed, 
i.e. no evaporation or antievaporation instabilities. 
In this case, thermodynamic proprieties were studied in Ref.\cite{Bamba:2012rv}. 

\section{Classical Evaporation and Antievaporation in non-diagonal tetrads}

Let us consider a non-diagonal tetrad base as follows: 
\be \label{e00}
e_{0}^{0}=e^{\rho},\,\,\,e_{3}^{3}=e_{0}^{1,2,3}=e_{1,2,3}^{0}=0\, ,
\ee
\be \label{e11}
e_{1}^{1}=\cos \psi \sin \theta \,e^{\rho}\, ,\,\,\,
e_{1}^{2}=\cos \psi \cos \theta \,e^{-\phi}\, ,\,\,\,
e_{1}^{3}=-\sin \psi \sin \theta \,e^{-\phi}\, ,
\ee
\be \label{e21}
e_{2}^{1}=\sin \psi \sin \theta \, e^{\rho}\, ,\,\,\,
e_{3}^{1}=\cos \theta \, e^{\rho}\, ,\,\,\,
e_{2}^{2}=\sin \psi \cos \theta \,e^{-\phi}\, , 
\ee
\be \label{e32}
e_{3}^{2}=\cos \psi \sin \theta \,e^{-\phi},\,\,\,
e_{2}^{3}=-\sin \theta \,e^{-\phi}	\, .
\ee

Under this ansatz, we obtain 
 \be \label{phi}
\delta \phi=A\sec \tau \cos(x-\bar{x})+B(\tan \tau)^{3/2}e^{\frac{1+2\cos^{2}\tau}{4\cos^{4}\tau}}\, , 
\ee
where $A,B$  are integration constants; 
corresponding to an horizon 
 \be \label{xh}
x_{h}=\bar{x}-\tau+{\rm arcsin}\left(\frac{\cos^{2}\tau}{A}\frac{d}{d\tau}\varphi(\tau) \right) \, , 
\ee
where 
 \be \label{varphi}
\varphi(\tau)=B(\tan \tau)^{3/2}e^{\frac{1+2\cos^{2}\tau}{4\cos^{4}\tau}}\, , 
\ee
Eq.(\ref{phi}) has a divergence in $\tau \rightarrow \pi/2$
--in the extreme time-like angle not included in the range of the Nariai solution. 
Depending on the integration constants, Eq.(\ref{xh}) solution 
is increasing or decreasing in time. 
The first class of instabilities corresponds to the classical antievaporation,
while the second class to the classical evaporation.

\section{No Bekenstein-Hawking in non-diagonal antievaporating Narai solution}

Let us consider the Raychaudhuri equation in $f(T)$-gravity \cite{Cai:2015emx}:
\be \label{theta}
\dot{\hat{\theta}}=-\frac{1}{3}\hat{\theta}^{2}-\hat{\sigma}_{\mu\nu}\hat{\sigma}^{\mu\nu}+\hat{\omega}_{\mu\nu}\hat{\omega}^{\mu\nu}-\mathcal{R}_{\mu\nu}U^{\mu}U^{\nu}
-\tilde{\nabla}_{\rho}\hat{a}^{\rho}-2U^{\nu}T_{\mu\nu}^{\sigma}\left(\frac{1}{3}h_{\sigma}^{\mu}\hat{\theta}+\hat{\sigma}_{\sigma}^{\mu}+\hat{\omega}_{\sigma}^{\mu}-U_{\sigma}\hat{a}^{\mu} \right)\,,
\ee
$\hat{\theta},\hat{\sigma},\hat{\omega},\hat{a}$ are
the expansion, shear, vorticity and acceleration in $f(T)$-gravity
and $\mathcal{R}_{\mu\nu}$ is the Ricci tensor corrected by contributions from the torsion: 
\be \label{RRRR}
\mathcal{R}_{\mu\nu}=R_{\mu\nu}-2\nabla_{\mu} T_{\rho}+\nabla_{\nu}K^{\nu}_{\mu\rho}+K_{\mu\lambda}^{\nu}K_{\nu\rho}^{\lambda}   \, ,
\ee
\be \label{RR}
R_{\mu\nu\rho}^{\sigma}=\partial_{\mu} \Gamma_{\nu\rho}^{\sigma} -\partial_{\nu} \Gamma_{\mu\rho}^{\sigma} +\Gamma_{\mu\lambda}^{\sigma}\Gamma^{\lambda}_{\nu\rho} -\Gamma_{\nu\lambda}^{\sigma} \Gamma_{\mu\rho}^{\lambda}    \, .
\ee
In general,  $\hat{\theta},\hat{\sigma},\hat{\omega},\hat{a}$ include the correction from the torsion as follows: 
\be \label{q1}
\hat{\theta}=\theta_{(GR)}-2T^{\rho}U_{\rho}\, ,
\ee
\be \label{q2}
\hat{\sigma}_{\mu\nu}=\sigma_{(GR)\mu\nu}+2h_{\mu}^{\rho}h_{\nu}^{\sigma}K_{(\rho \sigma)}^{\lambda}U_{\lambda}\, ,
\ee
\be \label{q3}
\hat{\omega}_{\mu\nu}=\omega_{(GR)\mu\nu}+2h_{\mu}^{\rho}h_{\nu}^{\sigma}K_{[\rho \sigma]}^{\lambda}U_{\lambda}\, ,
\ee
\be \label{q4}
\hat{a}_{\rho}=a_{\rho(GR)}+U^{\mu}K_{\mu\rho}^{\sigma}U_{\sigma}\, ,
\ee
$$\tilde{\nabla}_{\mu} U_{\nu}=\hat{\sigma}_{\mu\nu}+\frac{1}{3}h_{\mu\nu}\hat{\theta}+\hat{\omega}_{\mu\nu}-U_{\mu}\hat{a}_{\nu}\, ,$$
where 
$\dot{\hat{\theta}}=\frac{\partial}{\partial \lambda}\hat{\theta}$, where $\lambda$ is the affine parameter. 
In the optical null case, 
 $U^{a}\equiv k^{a}$ with $k^{a}=\frac{dx^{a}}{d\lambda}$, with $k^{2}=0$
 and 
$$\hat{\theta}=k^{a}_{;a}=2\frac{1}{\Sigma}\frac{d\Sigma}{d\lambda}\,.$$
We can define an emitting marginally outer 2-surface $\Sigma_{time-like}$
and the non-emitting inner 2-surface $\Sigma_{space-like}$. 

The marginally outer trapped 2-surface $\Sigma^{2d}_{MOT}$
has a topology of space-like 2-sphere with the condition
\be \label{theta}
\hat{\theta}_{+}(\Sigma_{MOT}^{2d})=0 \, ,
\ee
where $\hat{\theta}_{+}$ in a $S^{2}$-surface is 
the divergence of the outgoing null geodesics. 
The 
$\hat{\theta}_{+}$ decreases with the increasing of the gravitational field
-- for example $\hat{\theta}_{+}>0$ for $r>2M$ in the Schwarzschild case. 

The radius of the $S^{2}$-sphere $\Sigma_{MOT}^{2d}$
coincides with the Schwarzschild radius. 
$S^{2}$-spheres with smaller radii 
than $r_{S}=2M$ will be trapped surfaces (TS), i.e. 
$\theta(\Sigma^{2d}_{TS})<0$.
Such a topological definition can be 
generalized for 3d surfaces.
 The dynamical horizon 
is a marginally outer trapped 3d surface. 
It is foliated by marginally trapped 2d surfaces. 
In particular, a dynamical horizon can be 
foliated by a chosen family of $S^{2}$ 
with $\theta_{(n)}$ of a null normal vector $m_a$
vanishing while $\theta_{n\neq m}<0$,
 for each $S^{2}$. 
In particular, one can distinguish among 
an emitting marginally outer trapped 3d surface 
$\Sigma_{time-like}^{3d}$
and a non-emitting one 
$\Sigma_{space-like}^{3d}$
by their derivative of $\hat{\theta}_{m}$
 with respect to 
 an ingoing null tangent vector $n_{a}$:


\be \label{fds}
\hat{\theta}_{m}(\Sigma^{3d}_{time-like})=0,\,\,\,\,\,\,\frac{\partial \hat{\theta}_{m}(\Sigma^{3d}_{time-like})}{\partial n^{a}}>0
\ee
and  the non-emitting one is define as 
\be \label{sds}
\hat\theta_{m}(\Sigma^{3d}_{space-like})=0,\,\,\,\,\,\,\frac{\partial \hat{\theta}_{m}(\Sigma^{3d}_{space-like})}{\partial n^{a}}<0\, .
\ee
Now, adopting these definitions,  
we 
 demonstrate that the antievaporation will transmute  
the emitting marginally trapped 3d surface to 
a non-emitting space-like 3d surface. 
We can consider the Raychaudhuri equation associated to our problem.
Let us suppose an initial condition $\theta(\bar{\lambda})>0$
with $\bar{\lambda}$ an initial value of the affine parameter
$\lambda$. In the antievaporation phenomena, 
the null Raychauduri equation is bounded 
as 
\be \label{boundR}
\frac{d\hat{\theta}}{d\lambda}<-\mathcal{R}_{ab}k^{a}k^{b}\, ,
\ee
where $\mathcal{R}_{ab}k^{a}k^{b}$ is the effective contraction of the Ricci tensor (\ref{RRRR}) with null 4-vectors, corrected by torsion contributions, in turn governed by the EoMs (\ref{EOM}). 


Let us consider the antievaporation case:
for  $\lambda>\bar{\lambda}$, it is
$\mathcal{R}_{ab}k^{a}k^{b}>K>0$, where $K$ is the 0-th leading order of the scalar function $\mathcal{R}_{ab}k^{a}k^{b}(t)$.
Such a case coincide with the antievaporating  solutions by definitions.
So that
\be \label{rett}
\hat{\theta}(\lambda)<\hat{\theta}(\bar{\lambda})-K(\lambda-\bar{\lambda})-O\{(\lambda-\bar{\lambda})^{2}\}
\ee
Neglecting higher order terms, this leads to 
$\hat{\theta}(\lambda)<0$ for $\lambda>\bar{\lambda}+\hat{\theta}_{0}/K$, 
where $\bar{\lambda}$ and $\hat{\theta}_{0}\equiv \hat{\theta}(\bar{\lambda})$ are defined at a  characteristic time $\bar{t}$. 
For a small $\delta t$, 
a constant 0th contribution sourced by the torsion 
will cause an effective focusing term in the Raychauduri equation. 
So that, an emitting marginally trapped 3d surface will 
exponentially evolve to a non-emitting marginally one.

Now, let us consider a Bekenstein-Hawking pair in an antievaporating solution. 
They are imagined to be created in the black hole horizon 
as virtual pair. The external gravitational field can 
promote them to be real particles.
Then, a particle of this pair 
can tunnel outside the black hole horizon, 
with a certain characteristic time scale $\tau_{bh}$. 
With an understood correction to the Black hole entropy formula, this process seems compatible with Nariai solutions in diagonal tetrad choice. 
 Bekenstein-Hawking's calculations are performed in the limit of a static horizon 
an a black hole in thermal equilibrium with the environment.
This approximation cannot work for antievaporating black holes. 
In fact, the horizon is displacing outward 
the previous radius. 
Bekenstein-Hawking pair will be trapped in the black hole interior, 
foliated in space-like surfaces $\Sigma_{space-like}$.
But 
from a space-like surface,
the tunneling effect of a particle is impossible: 
otherwise causality will be violated. 
The suppression effect is a subleading effect if and only if the 
Bekenstein-Hawking radiation has a characteristic time 
$\tau_{bh}<\delta t$, where 
$\delta t$ is the minimal effective time scale in the external rest frame
for a
$\Sigma_{time-like}\rightarrow \Sigma_{space-like}$ transition. 
However, this cannot be possible for an arbitrary small $\delta t$. 
This leads to the conclusion that the Bekenstein-Hawking radiation is exponentially turned off in time. 
This conclusion is solid 
{\it for every antievaporating solution in $f(T)$-gravity} discussed in Section 4.
 

Now, let us comment what happen in the opposite case: evaporating solutions. 
In this case $f(T)$-gravity will source an extra anti-focalizing term in the the null Raychauduri equation.
This will cause exactly the opposite transition: a null-like horizon is pushed out the black hole radius and it will become 
time-like. Defining $\delta t$ as the transition time $\Sigma_{space-like}\rightarrow \Sigma_{time-like}$, 
Bekenstein-Hawking effect will happen if $\tau_{bh}<<\delta t$. 
However, with $\delta t<\tau_{bh}$, the Bekenstein-Hawking pair is pushed-off from the black hole horizon. 
In other words, they {\it both} will be emitted from the black hole. 
So that they can annihilate out-side the black hole producing radiation. 
Contrary to Bekenstein-Hawking radiation, unitarity is not violated in black hole formation during the gravitational collapse. 
In fact, firewall paradox is exactly coming by the entanglement of
the two pairs combined by the fact the one is falling inside the interior while its twin tunnels out. 
In our case, both are emitted out. 
In Bekenstein-Hawking case, outgoing information is exactly copied with the interior 
information. In our case, there is not any entaglement among black hole interior 
and external environment. 
So that this radiation does not introduce any new information paradoxes.

\section{Conclusions and outlooks}

In this paper,  extremal Schwarzschild-de Sitter solutions of  $f(T)$-gravity 
 have been  analyzed in quantum semiclassical regime. 
 These solutions are known in literature as Nariai Black Holes. 
 In $f(T)$-gravity, Nariai solutions are static in diagonal tetrad basis
 and their thermodynamic proprieties were studied in \cite{Bamba:2012rv}:
 a Bekenstein-Hawking radiation is emitted in this case \cite{Bekenstein,Hawking}. 
 However, the Nariai solution 
 has antievaporation or evaporation instabilities
 in non-diagonal tetrad basis \cite{Houndjo:2013qna}. 

In these two cases, we have studied the the null Raychauduri equation.
In the case of antievaporation, we have demonstrated that the torsion contribution provides 
a new term in Raychauduri equation 
trapping the emitting  surfaces in the black hole
space-like interior {\it before} the effective Bekenstein-Hawking  emission time.
Surprisingly, Bekenstein-Hawking radiation is suppressed. 
On the other hand, the evaporation instability, sourced by torsion contribution
of $f(T)$-theory, leads to a new peculiar quantum mechanical effect.
The external gravitational field will promote virtual pairs to a real particle-antiparticle in the null-like event horizon. 
However, before the Bekenstein-Hawking emission by tunneling, both particles are pushed out the black hole horizon.
In fact the torsion contribution provides an anti-focalizing term in the Raychauduri equation, 
pushing out space-like surfaces into time-like regions.   
As a consequence, {\it both} the  particle and the antiparticle are emitted by the black hole. 
This new radiation does not lead to paradoxical losing of unitarity in quantum mechanics.
The so called firewall paradox  is related to an entaglement of Bekenstein-Hawking pair, 
leading to an entanglement of the black hole interior and its external environment \cite{Braunstein:2009my,Almheiri:2012rt}. 
In our case, the entangled particles are both emitted out.  

The extension of the Einstein-Hilbert action as $f(R)$-gravity or $f(T)$-gravity
can be also motivated as EFT of quantum gravity. 
 An UV completion of $f(T)$-gravity is not still understood and in particular the quantization procedure in tetrad basis. 
 This is a crucial aspect to clarify in future. 
 Further other issues concern evaporation and antievaporation instabilities 
in virtual micro black holes pairs
  \footnote{The relevance of virtual black holes 
  contribution
  in vacuum energy density and in quantum information 
  processing was recently studied
  in Refs. 
  \cite{Addazi:2016jfq,Addazi:2017xur}.}. 
  In particular, it should be possible that 
  such instabilities promote the spontaneous 
  production of micro-black holes
  from virtuality. 
  Finally, 
  possible quantum chaos effects on the event horizon 
 may be crucially relevant in context of unstable antievaporating black holes
  \cite{Addazi:2015gna,Addazi:2016cad}.
  These aspects certainly deserve further investigations 
  beyond the purposes of this paper.


\begin{acknowledgments} 

I would also like to thank Salvatore Capozziello and Antonino Marciano for
useful discussions and remarks on this subject. 

\end{acknowledgments}

\end{document}